\documentclass[12pt,aps,prb,preprint,showpacs,showkeys]{revtex4}
\usepackage{graphicx}
\usepackage{amssymb}
\begin{document}
\title{Stratospheric Albedo Modification by Aerosol Injection}
\author{J. I. Katz}
\email[]{katz@wuphys.wustl.edu}
\affiliation{Department of Physics and McDonnell Center for the
Space Sciences \\ Washington University, St. Louis, Mo. 63130}
\thanks{Work performed under the auspicies of the Novim Foundation.  I thank
the Kavli Institute of Theoretical Physics at the University of California,
Santa Barbara, for hospitality.}
\date{\today}
\begin{abstract}
This paper reviews and develops the proposal, widely discussed but not
examined in detail, to use stratospheric aerosols to increase the Earth's
albedo to Solar radiation in order to control climate change.  The potential
of this method has been demonstrated by the ``natural experiments'' of
volcanic injection of sulfate aerosols into the stratosphere that led to
subsequent observed global cooling.  I consider several hygroscopic oxides as
possible aerosol materials in addition to oxides of sulfur.  Aerosol
chemistry, dispersion and transport have been the subject of little
study and are not understood, representing a significant scientific risk.
Even the optimal altitude of injection and aerosol size distribution are
poorly known.  Past attention focused on guns and airplanes as means of
lofting aerosols or their chemical precursors, but large sounding rockets
are cheap, energetically efficient, can be designed to inject aerosols at
any required altitude, and involve little technical risk.  Sophisticated,
mass-optimized ``engineered'' particles have been proposed as possible
aerosols, but the formidable problems of their production in quantity,
lofting and dispersion have not been addressed.
\end{abstract}
\pacs{92.30.Pq,92.70.Cp,92.70.Mn}
\keywords{geoengineering, stratospheric aerosols, climate modification}
\maketitle
\newpage
\section{Introduction}
Several historic volcanic eruptions (Tambora in 1815, preceding the
``Year [1816] without a Summer'' in the northeastern US, Krakatau in 
1883, El Chic\'on in 1982 and Pinatubo in 1991) were associated with
short-term ($\sim 1$--3 years) subsequent hemispheric cooling.  It
has been generally accepted for a long time that the eruptions caused
the cooling (and spectacular sunrises and sunsets) by injecting
aerosols into the troposphere and lower stratosphere, and that these
effects disappeared as the aerosols were removed from the atmosphere by
sedimentation or scavenging by hydrometeors.

All aerosols scatter sunlight, reducing the insolation at the surface,
and therefore cool the surface and the mixed boundary layer.  The
scattering, augmenting the Rayleigh scattering of clear air, makes
vivid sunsets.  Some aerosols (soot and some mineral dusts) also
absorb sunlight and heat their surrounding air, and indirectly the
ground, but this is believed to be a comparatively minor effect. 
These properties and effects of aerosols (unlike most of the rest of
global change and climatological research) are uncontroversial.

Several decades ago this led to the suggestion \cite{B74} that injection
of anthropogenic aerosols into the stratosphere could cool the climate,
were that desired.  The growing concern over global warming, together
with the expectation that it will increase with the increase in
anthropogenic greenhouse gases, particularly CO$_2$ but possibly also
CH$_4$ and perhaps others, has led to a revival of interest in the
injection of anthropogenic aerosols.  A recent extensive review was
presented by Rasch, {\it et al.\/} \cite{R08}, and a critical assessment by
Lacis, {\it et al.\/} \cite{L08}, who discuss in more detail issues of
particle agglomeration and scattering properties that we allude to here.

\section{Aerosol Scattering}

\begin{figure}
\centerline{[width=6in,height=6in]\includegraphics{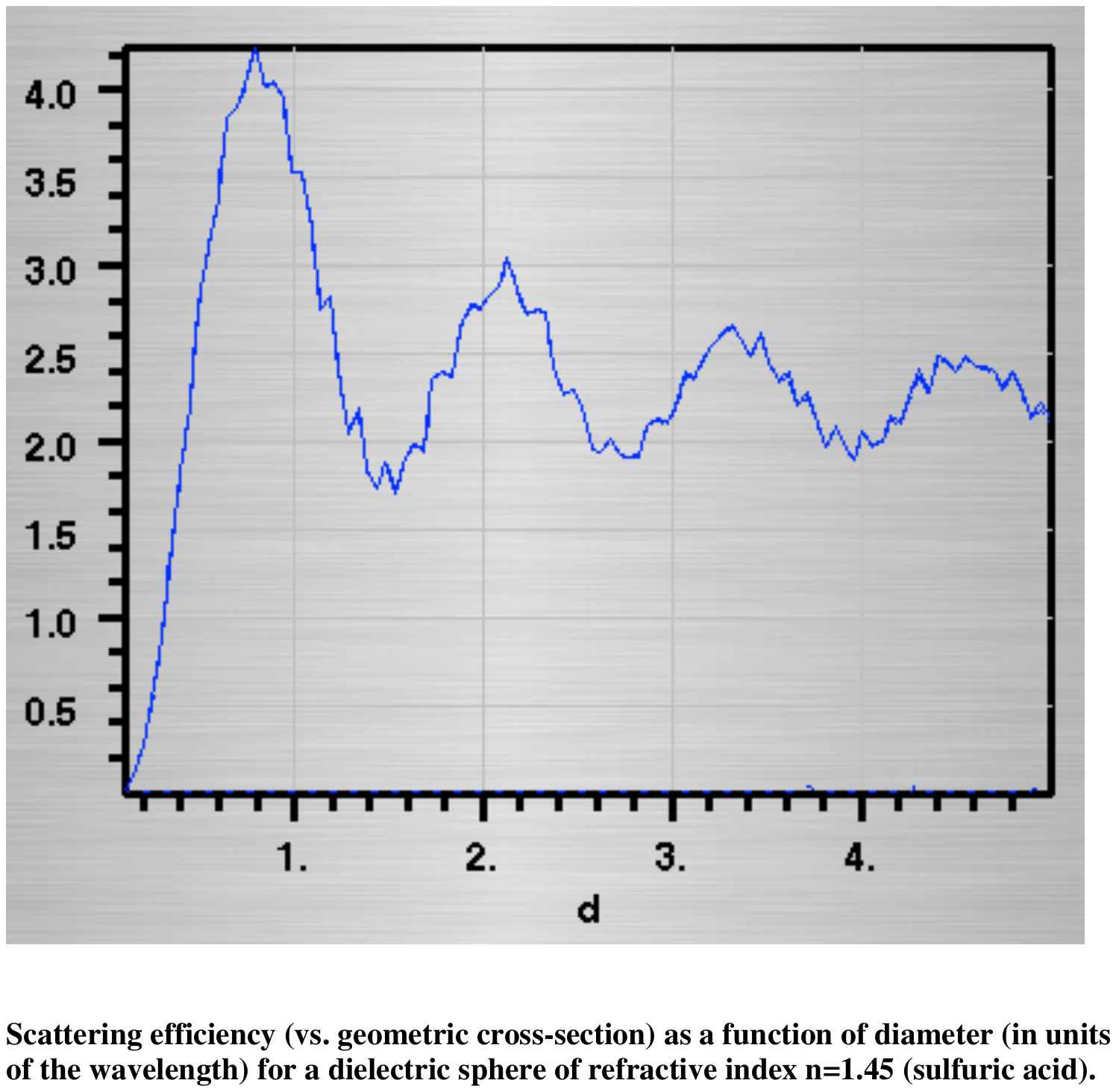}}
\caption{\label{miefig} Scattering efficiency, defined as total scattering
cross-section divided by projected area, {\it vs.\/} particle diameter $d$
in units of radiation wavelength, for dielectric spheres with $n = 1.42$
(sulfuric acid) \cite{MC08}.}
\end{figure}

The most efficient (per unit mass) spherical bulk-density scatterers
have radii of about 0.1 of the wavelength of the scattered radiation.
For the Solar spectrum, peaking (depending on whether the peak is per
unit wavelength or per unit frequency) on the red side of the visible
spectrum, this means radii $\approx 1000$ \AA.  Scattering from such
particles is described as Mie scattering, and results are widely
available.  Figure \ref{miefig} (\cite{MC08}) shows the scattering
efficiency as a function of the particle diameter for a sphere of
refractive index $n = 1.42$, characteristic of H$_2$SO$_4 \cdot
$H$_2$O\footnote{The equilibrium degree of hydration depends on the
activity (fugacity) of water vapor, equivalent to relative humidity
\cite{KPD08,W08}.  This particular composition is not a
stoichiometrically defined compound, but rather a representative
concentration, 84\% by mass of H$_2$SO$_4$, of sulfuric acid aerosol
in the stratosphere.}.

The asymptotic forms of the scattering cross-section are elementary.  For $d
\ll \lambda$ the cross-section is just the Rayleigh cross-section $\sigma
\propto d^6/\lambda^4$, leading to a mass efficiency $\epsilon \equiv
\sigma/(\pi \rho d^3/6) \propto d^3/\lambda^4$, so that very small particles
are completely ineffective.  We have retained the wavelength dependence
because it is also strong; particles that are effective in scattering blue
light are an order of magnitude less effective in scattering red light
(hence the blue sky, or blue diesel exhaust).  For $d \gtrsim \lambda$,
$\sigma \propto d^2$, (twice the geometric cross-section\footnote{For
macroscopic objects half the scattering is into forward angles ${\cal O}
(\lambda/(\pi d))$ and is not readily observed, so we are familiar with
the geometric cross-section only.}), so that $\epsilon \propto d^{-1}$, 
and large particles are also inefficient users of mass.

\section{Aerosol Materials}

Natural scattering aerosols are mostly soot (from fires), minerals (from
dust), salt (in the oceanic troposphere, from evaporated spray), water
(condensed from the atmosphere) and concentrated sulfuric acid (resulting
from oxidation of S to SO$_2$ to SO$_3$ followed by hydration).  Soot,
wind-lofted mineral dust and salt are generally limited to the troposphere,
where their lifetimes (against precipitation) are short.  Water is in
local equilibrium with its vapor (although often not in liquid-ice 
equilibrium; undercooled liquid drops are very common), and hydrometeors
are continually condensing and evaporating, so that any attempt to
increase the quantity in the stratosphere would probably be rapidly
redistributed through the vapor phase into larger ice crystals that
would precipitate.  Volcanic cooling is largely the result of mineral
and sulfate aerosols lofted into the stratosphere.  Their vapor pressures
are negligible (in the case of sulfuric acid, this refers to the vapor
pressure of the sulfur-containing SO$_3$) so the larger particles do not
grow at the expense of the smaller ones (except very slowly by
agglomeration), and lifetimes may be many years, depending on altitude.

For deliberate albedo modification we wish to use aerosols that scatter
sunlight with minimal absorption, minimal mass, minimal vapor pressure
(so they don't evaporate), are chemically stable in our oxidizing
atmosphere, even under the influence of the Solar UV flux at stratospheric
altitude, and minimal scattering or absorption of thermal infrared radiation
in the 8--14 $\mu$ window from the ground (this window is the vent in the
atmospheric greenhouse).  These criteria point to oxides of elements in the
third row of the periodic table (water is excluded by its high vapor
pressure) or of boron.   Transition metal oxides are excluded because they
are generally strong absorbers of visible light and second row elements
other than boron are excluded because they are either rare and toxic in all
chemical forms (beryllium) or greenhouse gases themselves.

Short ($\lambda/2$) fine (thickness and width $\sim 300$ \AA) aluminum wires,
coated with inert oxides of aluminum or silicon for protection against
oxidation, have been proposed \cite{TWH97} as mass-efficient scatterers,
either in the upper atmosphere or in space.  Although the mass efficiency is
spectacular, these wires must be mounted on much more massive sheets of
other materials (perhaps plastic) for handling.  In one space application
\cite{TWH97} the wires form a large two dimensional array whose form must
be maintained to achieve the desired diffraction pattern, requiring a stiff
and massive mount.  An area in excess of 1\% of the projected surface area
of the Earth must be covered with this material, either in space or in the
atmosphere.  As a rough guide to the thickness of cheap films that can be
manipulated, polyethylene (a very cheap plastic of low modulus and low
strength) garment bags are about 0.5 mil thick (one manufacturer \cite{CTS08}
quotes thicknesses of 0.43 and 0.65 mil), or more than 10 $\mu$.  Everyday
experience with these bags indicates that much thinner films of plastic are
likely to be difficult to handle.  Conventional high altitude scientific
balloons (zero overpressure) are made of 0.8 mil (20 $\mu$) polyethylene.
The mass of film required per dipole is likely to far exceed the mass of a
dielectric particle of similar scattering cross-section, which is typically
$\sim 0.01$ mil in diameter.

The remaining candidates (some of them are hydrated oxides and some will
hydrate further under stratospheric conditions) are 
\begin{enumerate}
\item Li$_2$O
\item B(OH)$_3$
\item Na$_2$O
\item MgO
\item Al$_2$O$_3$
\item SiO$_2$
\item H$_3$PO$_4$
\item H$_2$SO$_4$
\end{enumerate}
Sulfuric acid is most often considered because its precursor oxides are
the most abundant in volcanic aerosols (a consequence of the fact that,
uniquely in this list, they are volatile), and are a source of tropospheric
aerosols as a result of burning fuels with sulfur impurities.
\subsection{Choice}
We will focus on the oxides of boron, silicon, phosphorus and sulfur
(hydrated SO$_3$ is sulfuric acid).  The reason is that from the preceding
list these, and only these, have volatile hydrides that are expected to
oxidize to the oxides in the stratosphere.  Introduction as volatile
(gaseous or pressurized or cryogenic liquid) hydrides facilitates their
dispersion, minimizes coagulation by delaying oxidation until after they are
well-diluted in the stratosphere, and reduces the mass that must be lofted
because the oxygen is drawn from the air.

\section{\label{disp}Aerosol Dispersion}

The stratosphere is a difficult region to reach.  Before considering
the vehicle delivering the aerosol material (or its precursor) to the
stratosphere, we briefly discuss the problem of dispersion, for it
would be pointless to deposit a compact mass or large particles of material
that would be inefficient scatterers of light or that would immediately fall
out of the stratosphere.

It is difficult to grind a solid to the required particle sizes of
$\lesssim 0.1\,\mu$ so dispersion to these sizes much occur after release.
In addition, it is necessary that the particles be dispersed enough that
they do not rapidly reagglomerate.  This is the principal reason why we 
focus on materials that can be lofted as gaseous precursors that will not
agglomerate, and will produce potentially agglomerating particulates only
after they have been diluted to low density.

One may consider three possible means of dispersion: 
\subsection{High explosives}
High explosives may disperse a solid or liquid, but generally not
finely enough for our requirements.  Although the available shock energy
density (${\cal E}\gtrsim 10^{11}$ erg/cm$^3$) exceeds the energy required
to disperse individual atoms or molecules, even in a brittle solid nearly
all this energy is turned into bulk kinetic energy and shear rates are
much too low to fragment it to the required size.  A rough argument
comparing the energy in the shearing velocity field to surface energies
suggests that the characteristic fragment size will be
\begin{equation}\label{fragsize}
a \sim \left({\varpi R^2 \over {\cal E}}\right)^{1/3} \sim 0.01\ {\rm cm},
\end{equation}
where $R \sim 30$ cm is the overall size of the exploding system and $\varpi
\sim 10^2$ dyne/cm is the surface energy (surface tension) of the surfaces
that must be formed in dispersion.  In addition, the fragments are stopped
by the air in a distance $\sim R (\rho/\rho_a)^{1/3} \sim 50 R$, where
$\rho$ is the material density and $\rho_a$ is the density of the ambient
air.  This suggests that agglomeration of explosively dispersed solid or
liquid particles into sizes too large to be useful would be a serious
problem.
\subsection{Combustion}
Combustion of solids or liquids, perhaps initially dispersed by high
explosives, may produce aerosols.  Of the elements considered, Li, Na, Mg,
P, Al and S readily burn at atmospheric pressure (Al only if finely divided
or strongly heated).  An unresolved issue with the production of fine
aerosols by combustion of solids or liquids is that near the burning region
the particle density is very high and they may coagulate into sizes too
large for them to be mass-efficient scatterers.

It is not obvious that these elements would burn at stratospheric densities,
two or three orders of magnitude less.  Even if they do, agglomeration in
the immediate wake of combustion may be a serious problem that would require
experimental investigation.

The (very exothermic) burning of boron particles is problematic \cite{G06a}
even at full atmospheric density because of the tendency of their surfaces
to be covered with a tenacious coat of inert oxide.  For the same reason it
is essentially impossible to burn bulk silicon.  In fact, silicon is used in
high temperature micro-turbine engines \cite{E04} because of its resistance
to oxidation (and ease of lithographic fabrication).
\subsection{Slow Oxidation of Gases}
Slow oxidation following mixture into the air of gases containing the
desired cation in a reduced state may produce the desired oxide aerosols.
Because the reaction is comparatively slow, the gas may be widely dispersed
and mixed into the air at low concentration before it reacts, avoiding early
agglomeration of the particulates.  For this reason we restrict attention
to the elements boron, silicon and sulfur that have volatile compounds
(hydrides).

The use of H$_2$S as a source of sulfur has been widely proposed.  This
method may also work with boron (using diborane B$_2$H$_6$), phosphorus
(using phosphene PH$_3$) and silicon (using silane SiH$_4$).  The use of
these hydrides has the additional advantage (in comparison to making the
oxide on the ground and lofting it into the stratosphere) that the oxygen in
the oxide and any water of hydration, which represent 77\% of the mass of
B(OH)$_3$, 53\% of the mass of SiO$_2$, 56\% of the mass of P$_2$O$_5$ and
71\% of the mass of H$_2$SO$_4 \cdot$H$_2$O, need not be lofted with the
cation, but are taken without mass penalty from the surrounding air.
B(OH)$_3$ has the particular advantage that the molecular weight of boron is
about 1/3 of that of sulfur or phosphorus and 2/5 of that of silicon.

Hydrides could be lofted as gas (through a chimney, discussed later), as
rapidly evaporating cryogenic or pressurized liquids dispensed by aircraft
in the lower stratosphere, or as cryogenic or pressurized liquids contained
in artillery shells and dispersed (to droplets no smaller than estimated in
Eq.~\ref{fragsize}, and larger if small charges are used) in the
stratosphere by explosives.  The dispersed liquid drops will evaporate
rapidly and mix into the surrounding air, where the vapor will spread by
velocity shear and turbulent diffusion.

It is generally accepted that H$_2$S will oxidize rapidly to SO$_2$.  At
full (sea-level) atmospheric pressure B$_2$H$_6$, PH$_3$ and SiH$_4$ ignite
or explode readily in air.  Hence we are optimistic that oxidation will soon
follow evaporation, but the kinetics require detailed investigation, both
experimental and theoretical.

All of the candidate gases are quite toxic, but the oxides are all not
specifically toxic (although all, except for SiO$_2$, are harmful, chiefly
because of their acidity, if ingested in quantity).  All the gases are major
industrial chemicals, routinely handled without harm if proper precautions
are taken.  They could be manufactured on and lofted from remote uninhabited
islands (desirable, in any case, because most of the lofting mechanisms are
poor neighbors) at the required latitudes of injection.
\subsection{Choice}
For the reasons described, slow oxidation of a hydride precursor
(B$_2$H$_6$, SiH$_4$, PH$_3$ or H$_2$S) is probably the best method of
dispersing aerosols.  Of these materials, H$_2$S probably has the optimal
combination of lesser toxicity and high ratio of final aerosol mass to
lofted precursor mass.  Detailed investigation of the oxidation kinetics is
required to determine the size distribution of the final aerosols, which is
essential to the practicality of any aerosol method of albedo modification.

\section{Aerosol Agglomeration}

Agglomeration is the enemy of dispersion.  Particles that agglomerate
may become too large to be mass-efficient scatterers, and they may become
so large that gravitational settling removes them from the stratosphere
unacceptably rapidly.

The Brownian mean free path of a spherical particle of density $\rho$
and radius $r$ in air of density $\rho_a$ is
\begin{equation}
mfp_{part} = {4 \over 3 C_D}{\rho \over \rho_a} r \approx 5 \times 10^4 r,
\end{equation}
where $C_D \sim 1$ is its drag coefficient.  This result is valid in Knudsen
flow, in which the mean free path of the air molecules $mfp_a \gg r$, as 
is the case for particles of interest in the stratosphere.  When $mfp_{part}
\gg r$, as is always valid for such small particles, the coagulation time of
uncharged monodisperse particles
\begin{equation}
t_{coag} = {2 \over n K},
\end{equation}
where $n$ is the particle number density and the coagulation coefficient
\begin{equation}
K = A \sqrt{24 \pi k_B T r \over \rho} \approx 4 \times 10^{-9} A\ {\rm
cm^3/sec},
\end{equation}
where $A$ is the accommodation (sticking) coefficient and we have taken $\rho
= 1.8$ gm/cm$^3$ (sulfuric acid, but not far wrong for any of the materials
considered) and $r = 1000$ \AA; these results are similar to those of
\cite{R08}.
\subsection{Early Times}
Agglomeration may be a particularly acute problem in the initial stages of
dispersion when the particle density is high.  This problem is avoided if
the cation of the scattering material is introduced as its volatile hydride
so that particulates are not formed (because the oxidation kinetics are not
instantaneous) until the vapor has dispersed to low density.
\subsection{Late Times}
To provide 3\% albedo increase with particles with $r = 1000$ \AA\ and
$\sigma \sim 3 \times 10^{-10}$ cm$^2$ implies a column density $\sim
10^8$/cm$^2$.  Distributed over a column 10 km in height, $n \sim 10^2$
cm$^{-3}$, and $t_{coag} \sim 5 \times 10^6/A$ sec, or $2/A$ months.  
This is prohibitively short unless $A \lesssim 0.1$.  The observed
persistence of volcanic aerosols implies that this is the case, at least
for some of their components (not necessarily sulfuric acid).

We have not been able to do a literature search for empirical intrinsic
(uncharged) values of $A$ for the materials under consideration.  Coatings,
either monolayers acquired in the stratosphere or deliberately introduced
during production of particles, may keep $A$ sufficiently small. 
\subsection{Electrostatic Repulsion}
Electrostatic repulsion between like-charged particles is probably more
efficaceous.  A quantitative evaluation depends on knowing how the charge
is distributed on insulating solid particles, and on an energy
minimization calculation for conducting particles (such as sulfuric acid)
in which charge is redistributed as two particles approach.  A reasonable
rough estimate for particles each with net charge $Q$ is
\begin{equation}
A \approx \exp{\left(-f{Q^2 \over 2r k_B T}\right)},
\end{equation}
where the charge distribution and mobility factor $f$ is in the range
$1/2 < f \le 1$; $f = 1$ for spherically symmetric immobile charge
distributions.  Adopting this value, for $r = 1000$ \AA\ and $A = 0.1$ we
find $Q = 1.4 \times 10^{-9}$ esu, or three elementary charge units.  The
surface potential is then 0.04 Volt.

Particles are readily charged by friction if they are dispersed from a
dense cloud in which they collide frequently, but more than one composition
of particle must be present, with different electroaffinities, in order that
all or most of the same species have the same sign of charge.  This may
explain the charging of volcanic particles\footnote{As evidenced by the
lightning that accompanies volcanic eruptions.}, which have a variety of
compositions, but will probably not occur for chemically uniform
anthropogenic stratospheric aerosols.

In the stratosphere (particularly above much of the ozone layer) 
particles are likely to be charged by Solar ultraviolet photons, while
the electrons attach themselves to molecular species with positive
electron affinities. O$_2$ has an electron affinity EA$= 0.44$ eV but that
of N$_2$ is negative \cite{nist08}.  If the kinetics are rapid enough to
permit an approach to thermodynamic equilibrium, then the abundant O$_2$
molecules determine the chemical potential of the electrons.  The particles
will have a mean positive surface potential equaling the O$_2$ electron
affinity, and $Q \approx {\rm EA}\,r$.  Then
\begin{equation}
A \approx \exp{\left(-f{Q\,{\rm EA} \over k_B T}\right)} \approx \exp{\left(
f{{\rm EA}^2\,r \over e^2 k_B T}\right)} \approx \exp{(-538 f)} \lll 1.
\end{equation}
Under these conditions there would be no agglomeration.  The electrostatic
repulsion is a thermodynamic effect, not a kinetic one, so there is no
obvious reason why it should be reversed even in the absence of ionizing
ultraviolet radiation (nighttime).

The Arrhenius factor for thermal ionization of O$_2^-$ of $\exp{(-{\rm EA}/
k_B T)} \sim \exp{(5100^{\,\circ} {\rm K}/ T)}$ ranges from $\sim 10^{-10}$
at the cold (217$^{\,\circ}$K) tropopause to $\sim 10^{-8}$ at the warmer
(271$^{\,\circ}$K) stratopause, so we may expect the electrons to remain
bound to O$_2$ molecules.  Neutralization may occur by aerosol-O$_2^-$
collisions, at a rate that must be calculated.  The lower the density of
aerosols, the fewer O$_2^-$ and the slower the neutralization, so this
process may set an upper bound to the nighttime degree of aerosol
ionization.  However, nonequilibrium chemical kinetics is complicated and
may lead to surprises.

\section{Aerosol Lofting}

A number of methods of lofting particulates (or, more likely, their
chemical precursors, for reasons discussed in the previous section) to
stratospheric altitudes may be considered.  The
feasibility of each depends on the material being lofted and on the means
of dispersion, so these three lists are not independent.
\subsection{Aircraft}
Injection of megaton quantities of sulfuric acid precursors into the
stratosphere requires heavy lift aircraft that can fly at these altitudes.
The KC135 has a service ceiling of 50,000$^\prime$ (17 km), at the lower
edge of the equatorial stratosphere.  The USAF has an inventory of several
hundred, 161 of which (along with 38 similarly capable B52s) are scheduled
for retirement in the next few years.  Gaskill \cite{G06} has suggested
their use to introduce aerosols or their precursors into the stratosphere.
Other heavy lift aircraft, such as commercial airliners, are not capable
of reaching these altitudes.

The chief limitation of the KC135 and B52 is their altitude.  It may be
that to obtain sufficiently long aerosol residence times altitudes of
30--50 km are required.  These altitudes are essentially unreachable by
aircraft.  The presence of equatorial stratospheric upwelling suggests that
these higher altitudes may not be necessary, but only a detailed transport
calculation, founded on empirical velocity field and turbulent transport
data, combined with tracer experiments, can answer this question. 
\subsection{Guns}
Naval guns have been proposed as a means of stratospheric
lofting.  The summit of the naval artillery art was achieved in the Iowa
class battleships, whose $50 \times 16^{\prime\prime}$ guns\footnote{The
first number is the barrel length in units of the caliber.} fired a 2700
lb shell with a muzzle velocity of 2500 ft/sec (762 m/sec) at a firing rate
of two shells per minute \cite{iowa08}.  In the absence of an atmosphere,
such a shell fired upwards would reach an altitude of 29.6 km.

Air drag reduces the muzzle velocity by a multiplicative factor
\begin{equation}
{v_f \over v_i} = \exp{\left[-{C_D \over 2}\left({1000\ {\rm gm/cm^2} \over
BC}\right)\right]} \approx 0.90,
\end{equation}
where we have taken a mean drag coefficient $C_D = 0.2$, used the ballistic
coefficient of the shell $BC = 944$ gm/cm$^2$ and made the approximation
that the atmospheric scale height is small compared to the altitude reached.
Drag then reduces the attainable altitude to 24.0 km.  This may be adequate
for injection into the upwelling lower equatorial stratosphere, but would
lead to short residence times in the downwelling polar stratosphere.

The main gun on the M1 tank has a muzzle velocity up to 1.7 km/sec,
illustrating what can be achieved with conventional chemical propulsion.
This might na\"ively suggest a maximum altitude of 92 km (taking the same
value of $BC$).  However, even when scaled to $16^{\prime\prime}$ (406 mm)
caliber from 120 mm, this high muzzle velocity is only achieved with a much
lighter projectile and a ballistic coefficient less by a factor of 3--7,
depending on which of the variants of the M829 ammunition is used for
comparison \cite{m829}.  For example, the 10 kg M829A3 round has a muzzle
velocity of 1.555 km/sec.  Scaled to 406 mm, it would have $BC = 299$
gm/cm$^2$.  Because air drag would be more important, it could reach an
altitude of 63 km (rather than 92 km) with a total shell mass of 387 kg, of
which 300 kg might be payload.  Optimal design of a gun for stratospheric
injection to 30--50 km requires engineering tradeoffs among these factors,
but these altitudes are clearly achievable.

With the demonstrated firing rate of two rounds per minute for
$16^{\prime\prime}$ naval guns, a nominal 1 MT/yr injection rate would
require three guns firing at their maximum rate.  Tank guns have a firing
rate roughly ten times higher, but it is unclear how this would scale to
$16^{\prime\prime}$ caliber, at which the scaled tank round would carry
1/3 the payload of the naval gun.  The optimal gun would lie somewhere
between these limits, depending on the desired altitude of injection.

The barrels of tank guns must be replaced every few hundred rounds because
of erosion, but this is likely to be less frequent at the lower muzzle
velocity we require.  The chief consequence of barrel erosion is reduced
accuracy, which is not an issue for geoengineering, so it is probably safe
to assume that each barrel is capable of firing 1000 rounds, and possibly
many more.  A nominal 1 MT/yr system with 300 kg payloads per round would
fire about 3,000,000 rounds/yr from a total of three guns, and would consume
no more than 3000 barrels/yr, corresponding to three barrel replacements
daily per gun, and perhaps many fewer.  

The cost of the shells (not including the payload) may be $\sim
{\cal O}$(\$10,000).   This is only a guess as to the cost of these very
simple shells, and is meant to be conservative.  For comparison, the
technically sophisticated JDAM guidance package is estimated to cost
\$20,000 per item \cite{jdam}.  Mass production would reduce the unit cost
far below that typical of low production-run peacetime military systems.
This nominal cost corresponds to a lofting cost of \$30/kg, or \$30
billion/MT.  More massive shells with parameters closer to those of the
extant $16^{\prime\prime}$ naval guns would probably have a lower cost per
unit mass lofted.  The guns, barrels, and other components are likely to be
a small fraction of the cost of the shells because of the economies of
using them continually over a long period.  Unlike military systems, no
elaborate turret capable of aiming accurately over wide angles would be
required; the geoengineering guns simply point up.
\subsubsection{Davis guns}
Davis guns are (in principle) recoilless; if their barrel is rifled they
are also known as recoilless rifles.  The absence of recoil is achieved
with a barrel open at both ends, with a wad of soft material expelled out
the back to take up the recoil of the projectile.  They were first developed
by CMDR Davis, U.~S.~Navy, in 1912--14 in order to solve the problem of
mounting a large-caliber gun on a fragile airplane.  A modern
$12^{\prime\prime}$ Davis gun operated for many years at the Tonopah test
range, firing downward to study earth penetration.  Because of their small
recoil, Davis guns are capable of firing massive projectiles without
enormously robust and expensive mounts, but because of the recoil mass they
are energetically inefficient and unlikely to be able to loft material into
the stratosphere.
\subsection{Rockets}
The required velocities 1--1.5 km/sec are a fraction of the exhaust velocity
$v_e$ (usually parametrized as specific impulse ISP$\equiv v_e/g \approx
300$ s) of solid rocket fuels.  The payload mass fraction $1 -
\exp{-v_b/v_e} \approx 0.6$--0.7, where $v_b$ is the rocket velocity at fuel
burnout (exhaustion), of the launch weight.  This is greater than for high
velocity tank rounds, in which the propellant mass exceeds the projectile
mass, but probably less than that of naval gun rounds.

Air drag is less important for rockets than for guns because rockets may be
slenderer than gun-launched projectiles, increasing their $BC$, and because
the burnout velocity of a rocket is only achieved above the denser parts of
the atmosphere.  Additional advantages of rockets include a milder launch
environment, permitting the payload to be carried in a thin-walled vessel
rather than than a massive artillery shell, and the absence of a massive
breach and barrel to contain the confined burning propellant.  We cannot
make a quantitative estimate, but expect the cost of rocket lofting to be
substantially less than that of guns\footnote{Given these arguments, why have
militaries mostly used guns, except at very long ranges?  The reasons are:
1.~Rockets require sophisticated technology to guide them to a target, while
guns are simply pointed (allowing for gravity, air drag and wind).  2.~A
gun-launched projectile has its full velocity out of the muzzle, while a
rocket may not have a lethal impact, and may even be so slow as to be
avoidable by the target, until a substantial distance from its launch.  None
of these arguments apply to geoengineering}. 
\subsection{\label{chimney}Balloon-Supported Chimneys}
It may be possible to inject gases into the stratosphere through a tube
suspended from a balloon\footnote{A tube would be required because if the
energy injected in the tropopause's Brunt-Vis\"al\"a time ($\sim 30$ sec) is
less than tens of megatons, any unconfined tropospheric injection would mix
in the troposphere and not reach the stratosphere.  Volcanoes are
sufficiently energetic to avoid this, but continual injection is not.}.   
Such a tube resembles a very high chimney, but must be suspended from a
a balloon at its upper end.  The buoyancy of the balloon must be sufficient
to support the weight of the tube.  It must also prevent tropospheric winds
from turning the tube horizontal, thereby pulling its upper end below the
required injection altitude.  This section contains some very rough
estimates, and is no substitute for an engineering design study.
\subsubsection{Chimney materials}
The materials considered have been high molecular
weight polymers such as Spectra, liquid crystal polymers such as Vectran,
and aramids such as Kevlar.  These materials all have uniaxial (aligned
fiber) strength $\approx 30$ KBar, Young's modulus $\sim 1$ Mbar and
density 1--1.5 gm/cm$^3$ \cite{FS04}.  If the fibers are distributed
orthogonally in a thin sheet its biaxial strength and modulus along the
axes may approach half the uniaxial value, but it may be very weak in
diagonal tension in which fibers can slide over one another.  If the
fibers are isotropically distributed very few will be aligned in any
direction and the material's tensile strength will depend on their
resistance to sliding, not on their uniaxial tensile strength.  The loads
on the tube will be predominantly along its length, so its fibers may be
oriented in that manner.  However, it is unclear what tensile strength to
assume for a balloon fabric in isotropic plane tension.  We parametrize the
strength by $S \equiv Strength$/(10 Kbar).  For the tube $S \sim 1$ may be
reasonable, but for the balloon $S$ may be much smaller.
\subsubsection{Flow through the tube}
Using standard methods \cite{M05} we have made a rough estimate of the
tube diameter required to accommodate the nominal 1 MT/yr ($3 \times 10^4$
gm/sec) flow through the tube, assuming a driving pressure roughly
comparable to ambient, as will be driven by buoyancy if the molecular
weight of the gas is a fraction of that of air.  This condition is 
satisfied if the hydride gas is diluted two- or three-fold with hydrogen.
The result is a radius at the upper end of the tube, assuming a pressure
there of 30 mbar (about 30 km altitude), to deliver $3 \times 10^4$ gm/sec
($5 \times 10^7$ m$^3$/day) of $r_t \approx 2$ m.  The lower end may be
several times narrower because of the higher density there.
\subsubsection{Aerodynamic loads}
The tube passes through the troposphere into the stratosphere, and will
occasionally encounter the jet stream.  The aerodynamic load for a length
$L$ immersed in a jet stream of speed $v = 50$ m/sec is
\begin{equation}
F = C_D L v^2 r_t \rho_a \sim 5 \times 10^{11}\ {\rm dyne},
\end{equation}
where we have taken $L = 1$ km as the jet stream depth, $C_D = 1$ and
$\rho_a = 1 \times 10^{-3}$ gm/cm$^3$ for the upper troposphere.  It may
be possible to reduce this load by a factor 2--3 if the tube is 
aerodynamically shaped with a ``weather vane'' to turn it into the wind.
In addition, we have made the very conservative assumption that the tube
has a constant diameter.  In fact, the portion at jet stream altitudes
may be a few times narrower because of the higher density there (we have
tacitly assumed the gas in the tube to be in pressure equilibrium with
the ambient air), reducing the aerodynamic load in proportion.

The cross-section required to bear this load in tension is 50 $C_D/S$
cm$^2$.  The weight of a tube of length $L_t = 50$ km is then
\begin{equation}
W = gM = {g \rho_t L_t C_D \over S}\ 4 \times 10^{11}\ {\rm dynes}.
\end{equation}

Of course, the previous calculation is not self-consistent.  The tube
must bear its own weight as well as that of any aerodynamic load.  We
could solve the self-consistent equation, but instead make the following
qualitative points:
\begin{enumerate}
\item The balloon must support a load $\sim 10^{12}$ dyne.
\item It is essential that the along-axis tensile strength of the tube
material be ${\cal O}$(10) Kbar.
\item Minimizing $C_D$ by aerodynamic shaping and optimal orientation of
the tube has large benefits.
\end{enumerate}
We conclude that the chimney may be feasible, but involves significant
technical risks in material and aerodynamic performance.

\subsection{Photophoresis}

Photophoresis (lift provided by asymmetric surface properties of a small
oriented particle not in thermal equilibrium with its gaseous environment)
has been suggested as an explanation of the presence of tropospheric soot in
the stratosphere \cite{R96,P00}.  Keith \cite{K08} has suggested its
application to deliberately engineered aerosols for the purpose of 
geoengineering.  Photophoresis requires a significant temperature difference
between the particle and the gas, particles of low density, and particles
with an asymmetric thermal accommodation coefficient and an offset between
their center of mass and center of drag.

Pueschel, {\it et al.\/} \cite{P00} found that ``fluffy'' soot aggregates
with mean densities of a few tenths of a gm/cm$^3$ could, if sufficiently
asymmetric, have sufficient diurnally averaged photophoretic lift to
overcome gravity.  However, the particles required to increase the Earth's
albedo must not (unlike soot) absorb a significant amount of Solar
radiation, and are expected to have higher densities (1.8 gm/cm$^3$ for
sulfuric acid, and somewhat greater for other materials).  Sulfuric acid
aerosols would be spherical liquid drops without any surface asymmetry or
offset between their centers of gravity and of drag.  Net photophoretic lift
appears unlikely for aerosols produced by the processes of Section
\ref{disp}.

Carefully engineered particles \cite{K08} might do much better.  However,
their scattering properties are not likely to be a great improvement over
those of mineral or liquid particles of similar dimensions, so for them to
be useful it must be possible to fabricate and disperse them in megaton
quantities at reasonable cost.
\subsection{Choice of Lofting Mechanism}
Two of the lofting concepts considered, rockets and guns, are technically
mature and would only require engineering development.  Rockets may be
substantially cheaper.  Chimneys would require extensive research and
development, and it is difficult to estimate their cost.  Photophoresis 
raises major questions of the ability to engineer and mass-produce suitable
particles that do not absorb visible light but have sufficient photophoretic
lift to loft them into the stratosphere.
\section{Balloons}
Balloons have at least three potential applications in geoengineering.
\begin{enumerate}
\item A means of carrying material to stratospheric altitude.  The material
might be a gas (such as a hydride precursor of aerosols) filling the
balloon itself, or in a vessel hanging from the balloon.
\item A source of lift for a chimney, as discussed in \ref{chimney}.
\item As reflective objects that themselves modify the Earth's albedo..
\end{enumerate}
These three applications require different kinds of balloons meeting
different technical criteria.  They are best considered according to their
required lifetimes rather than according to their application.

\subsection{Short-lived balloons}

The buoyancy of a balloon in pressure equilibrium with the ambient air is
\begin{equation}
B = g (M_a - M_b) = g M_a \left(a - {\mu_b \over \mu_a}
{T_a \over T_b}\right),    
\end{equation}
where $M_a$, $\mu_a$ and $T_a$ are the mass, molecular weight and
temperature of the displaced air and $M_b$, $\mu_b$ and $T_b$ are the mass,
molecular weight and temperature of the gas filling the balloon.  Because
if $T_a = T_b$ $B$ is independent of altitude (equivalently, independent of
atmospheric pressure) such a balloon has no equilibrium height.  If $B$
exceeds the load it will rise indefinitely, until (if open at the bottom) it
spills lifting gas, or (if closed) it bursts from internal overpressure once
the skin expands to its maximum volume.  If $B$ is less than the load it
sinks to the surface of the Earth.

In practice, the altitude of a pressure-equilibrium balloon, such as those
used to loft scientific payloads to the stratosphere, is controlled by
dumping ballast.  If the filling gas were always in thermal equilibrium with
the air it would remain at a constant altitude indefinitely, once enough
ballast had been dumped (or gas spilled) that the lift equals the load.  But
because of the diurnal variation in Solar heating of the skin (and advective
heat transport to the filling gas) $T_a/T_b$ varies and ballast or gas must
be expended daily.  As a result, pressure-equilibrium balloons have flight
durations of $\cal O$(10) days, except during polar summer and winter
(``midnight Sun'' or ``noontime night'') when there is no diurnal Solar
heating cycle.  Somewhat longer durations may be obtained if they are made
of material that is less absorbing of Solar near-infrared radiation than
polyethylene, in order to reduce the magnitude of their diurnal temperature
swings, if they are baffled inside to reduce transport of heat from the
skin, or if they are aluminized to reflect sunlight. 

These balloons must be in pressure equilibrium with the ambient air because
they are made of very weak material (typically 0.8 mil polyethylene, with a
tensile strength of $\sim 300$ bars and even lower yield threshold, so that
a 100 m radius balloon begins plastic flow at an overpressure of $\lesssim
100$ dyne/cm$^2$, about 10$^{-4}$ bar or about $10^{-2}$ of a float pressure
of 10 mbar at about $120000^\prime$.  Such a balloon is very cheap and
light, and they have been used to loft scientific payloads for many years.
\subsubsection{Delivery vehicles}
Short-lived balloons are satisfactory if the goal is only to deliver
materials to the stratosphere.  Gaseous material may either be mixed with
hydrogen or helium as the lifting gas, or (if liquid or solid) may be
suspended from the balloon, as are scientific payloads.  Volatile materials
(such as the precursor hydrides we have considered) are better carried
as gases to take advantage of their buoyancy which, at least partially,
offsets their weight.  This also avoids the need to lift the parasitic
weight of a cryogenic or pressurized container.

Balloon delivery of materials has been considered and rejected on the
grounds that the number of balloons is excessive and that the large number
of expended balloons falling to the Earth would pose an unacceptable risk to
the environment \cite{R08}.  Balloons that failed to vent or burst in the
planned location might also pose a risk to aviation upon their unpredictable
descent.  These objections are difficult to evaluate.

A typical scientific balloon operation may cost several hundred thousand
dollars, and lofts a payload of order a ton, suggesting a cost per unit mass
perhaps 1--10 times that of artillery lofting.  The launch of such a balloon
is a tricky operation that depends on favorable weather (low wind) at the
launch site.

\subsection{Long-lived balloons}

If we wish to use a balloon to support a chimney, or to effect a long-term
reduction in the Earth's albedo, we must avoid daily expenditure of lift gas
or ballast.  The solution to this problem is an overpressure balloon, whose
volume is essentially independent of its temperature.  The concept is old,
but its realization has depended on the development of better materials
\cite{cnes,hatb}.
\subsubsection{Lift}
To provide $10^{12}$ dynes of lift at the 30 mbar
level requires a volume of about $2 \times 10^{13}$ cm$^3$, or a radius
$r_b \approx 170$ m.  The overpressure $\Delta P$ it must support is a
fraction $f_{var}$ (the fractional diurnal temperature variation)
of ambient, or $\sim 10^4$ dyne/cm$^2$.  The required wall thickness is 
\begin{equation}
\Delta r_b \ge {f_{var} P \over 2 S} r_b \approx {0.01 \over S}\ {\rm cm}.
\end{equation}
The ratio of the weight $W_{skin}$ of its skin to its buoyant lift is then
\begin{equation}
{W_{skin} \over B} = {3 \over 2} {f_{var} P \over 10^{10} S} {\rho_{skin}
\over \rho_a (1 - \mu_b/\mu_a)} \approx 0.05/S,
\end{equation}
where we have taken a temperature of 250$^{\,\circ}$K, $\rho_{skin} = 1.5$
gm/cm$^3$, $f_{var} = 0.3$ and H$_2$ as the filling gas.

The importance of the material strength is evident.  For example, Mylar has
an ultimate tensile strength of 1.5 Kbar \cite{KMM65}, which gives $W_{skin}
/ B \approx 0.3$, so the requirement to contain the overpressure of large
temperature swings would exact a large price in a Mylar balloon's lifting
capability.  The materials discussed in \ref{chimney} are much stronger, but
their behavior when used to make membranes subject to isotropic tension must
be understood.
\subsubsection{Baloon Albedo}
It is also possible to consider using balloons themselves to increase the
albedo of the Earth if they are coated with a material with high
reflectivity \cite{TWH97}.  The obvious choice is vapor-deposited aluminum,
Aluminized plastic films are widely used in applications ranging from 
insulation (where the aluminum layer inhibits radiative transport of heat)
to space flight.  A layer of $\sim 300$ \AA\ of aluminum is sufficient to
reflect most incident Solar visible and near-infrared radiation, while
transmitting most of the upwelling mid-infrared radiation of the Earth.

The minimum size of such a balloon is set by the requirement that the 
aluminum, which contributes negligibly to its strength, have a weight small
compared to that of the underlying plastic.  This implies $\Delta r \gtrsim
0.3 \mu$.  The thinnest plastic films of which we are aware are $0.9\,\mu$  
thick.  The films used in Solar sail experiments have been either $5\,\mu$
Mylar or $7.5\,\mu$ Kapton.  We do not know how thin films can be made from
the high-strength materials discussed in Section \ref{chimney}.  For $\Delta
P = 10^4$ dyne/cm$^2$, $r = 2 \times 10^6 S = {\Delta r \over 1\,\mu} S 
\times 200 \ {\rm cm}$, assuming the balloon is designed to minimize
$W_{skin}/B$.  Smaller balloons are possible (letting $S$ be the maximum
achieved tensile stress in the skin, rather than the material's limiting
stress), but the minimum radius for which the buoyancy is positive is about
5--10 cm.

Unless there is a breakthrough in making and handling ultra-thin films, the
minimum diameter of an overpressure balloon will be $\sim$ 10--20 cm.  Its
mass would be $\sim 0.5$ gm.  Although very light, it would be large and
very strong.  A rain of such balloons that have reached the end of their
lives would be a significant hazard to wildlife and conceivably to aviation.
\section{Research Program}
This report has discussed a number of questions involving aerosol properties
(entirely apart from our understanding of climate, either natural or subject
to anthropogenic forcing) that must be answered before it can be known if
aerosol mitigation of the thermal effects of increasing atmospheric CO$_2$
is feasible.  We list a number of issues in basic science, each of which
needs both theoretical and experimental investigation:
\begin{enumerate}
\item Chemical kinetics of oxidation of gaseous hydride precursors
\item Physical kinetics of aerosol aggregation
\item Aerodynamics and aeroelasticity of chimneys and balloons
\item Properties of candidate chimney and balloon materials
\item Stratospheric transport aerosols
\begin{enumerate}
\item Wind fields
\item Turbulent diffusion
\item Photophoresis
\item Sedimentation
\end{enumerate}
\item Engineered aerosols
\item Side effects of anthropogenic stratospheric aerosols
\begin{enumerate}
\item Stratospheric chemistry (ozone depletion?, {\it etc.\/})
\item Ecological consequences of increased diffuse (scattered) radiation flux
\item Tropospheric and terrestrial effects of precipitated aerosols
\end{enumerate}
\end{enumerate}

In addition, there are many engineering design issues that must be
addressed before any aerosol climate modification plan can be developed.
We believe the basic science questions should be answered first, so that
the engineering efforts can be directed in a most productive manner.

\end{document}